\documentclass[%
 reprint,
superscriptaddress,
frontmatterverbose, 
nofootinbib,
nobibnotes,
 amsmath,amssymb,
 aps,
prb,
floatfix,
longbibliography]{revtex4-2}
\usepackage{graphicx}
\usepackage{dcolumn}
\usepackage{bm}
\usepackage{hyperref}
\hypersetup{colorlinks = true, citecolor = blue, linkcolor=blue, urlcolor=blue, pdfauthor=author}
\begin{document}
\preprint{APS/123-QED}

\title{Polarization State Conversion through Chiral Butterfly Meta-Structure}

\author{Hicham Mangach}
\affiliation{%
Light, Nanomaterials Nanotechnologies (L2n), CNRS-ERL 7004, Université de Technologie de Troyes,10000 Troyes, France\\
}%

\affiliation{%
Laboratory of optics, information processing, Mechanics, Energetics and Electronics, Department of Physics, Moulay Ismail University, B.P. 11201, Zitoune, Meknes, Morocco\\
}%

\author{Younes Achaoui}
\affiliation{%
Laboratory of optics, information processing, Mechanics, Energetics and Electronics, Department of Physics, Moulay Ismail University, B.P. 11201, Zitoune, Meknes, Morocco\\
}%

\author{Muamer Kadic}
\affiliation{%
Université de Franche-Comté, Institut FEMTO-ST, CNRS, 25000 Besançon, France\\
}%

\author{Abdenbi Bouzid}
\affiliation{%
Laboratory of optics, information processing, Mechanics, Energetics and Electronics, Department of Physics, Moulay Ismail University, B.P. 11201, Zitoune, Meknes, Morocco\\
}%

\author{Sébastien Guenneau}
\affiliation{%
UMI 2004 Abraham de Moivre-CNRS, Imperial College London, SW7 2AZ, UK\\
}%
\affiliation{%
The Blackett laboratory, Physics Department,  Imperial College London, SW7 2AZ, UK\\
}%

\author{Shuwen Zeng }
\email{{Corresponding author: shuwen.zeng@cnrs.fr}}
\affiliation{%
Light, Nanomaterials Nanotechnologies (L2n), CNRS-ERL 7004, Université de Technologie de Troyes,10000 Troyes, France\\
}%

\date{\today}

\begin{abstract}
The recent emergence of chirality in mechanical metamaterials has revolutionized the field, enabling achievements in wave propagation and polarization control. Despite being an intrinsic feature of some molecules and ubiquitous in our surroundings, the incorporation of chirality into mechanical systems has only gained widespread recognition in the last few years. The extra degrees of freedom induced by chirality has propelled the study of systems to new heights, leading to a better understanding of the physical laws governing these systems. In this study, we present a structural design of a butterfly meta-structure that exploits the chiral effect to create a 3D chiral butterfly capable of inducing a rotation of 90° in the plane of polarization, enabling a switch between various polarization states within a solid material. Furthermore, our numerical investigation using Finite Element Analysis (FEA) has revealed an unexpected conversion of compressional movement to transverse movement within these structures, further highlighting the transformative potential of chirality in mechanical metamaterials. Thus, revealing an additional degree of freedom that can be manipulated, namely the polarization state.
\end{abstract}
                            
\maketitle


\section{\label{sec:level1} Introduction}

\begin{figure}[!htbp]
\centering
\includegraphics[width= 7.5 cm]{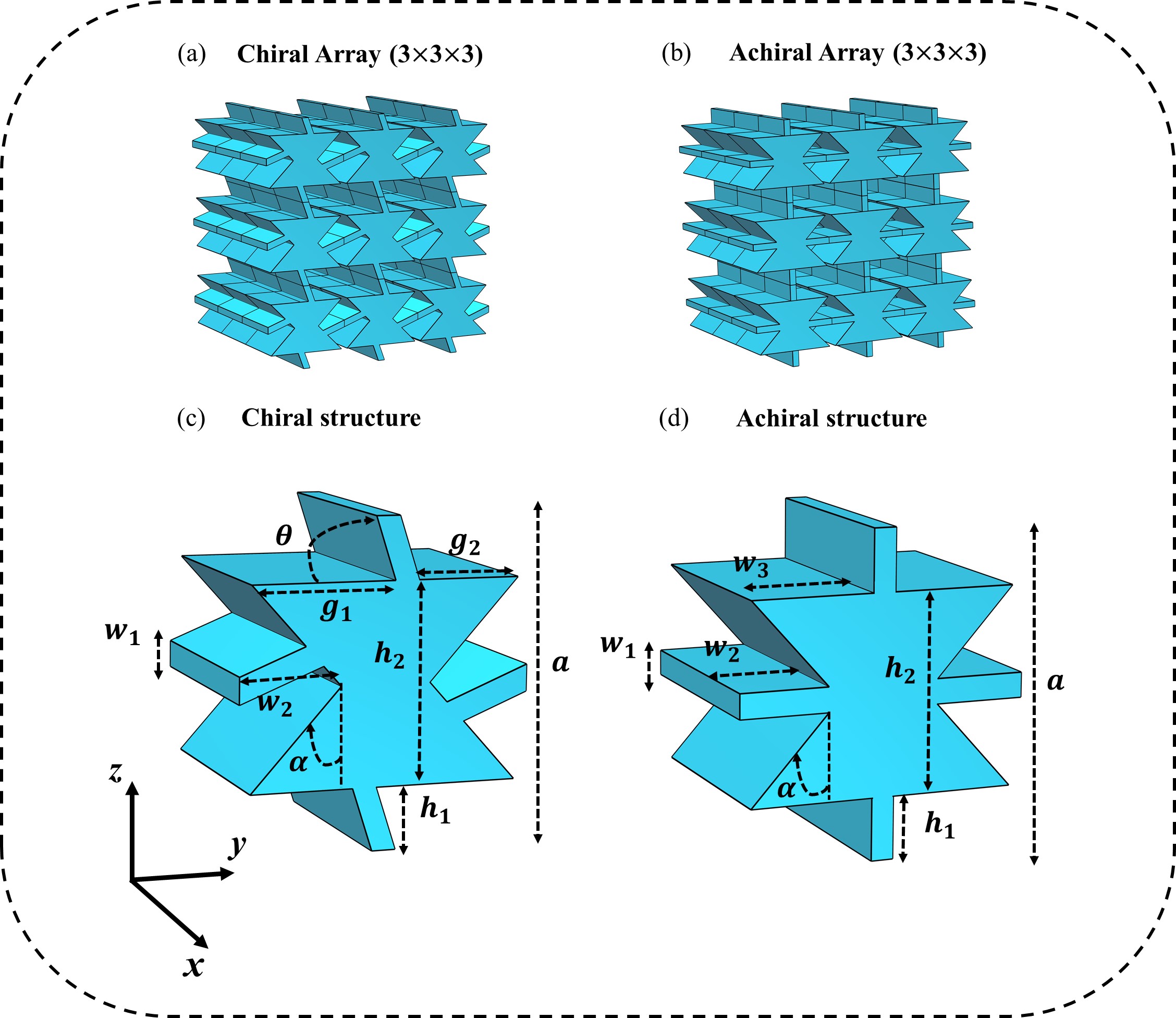}
\caption{\label{fig:01} (a) and (b) the 3D arrays encompassing chiral and achiral butterfly meta-atoms, respectively, with a unit cell dimension of (3 $\times$ 3 $\times$ 3). (c) and (d) The 3D schematic design of chiral and achiral butterfly unit-cells with a uniform spatial period of $a$ in all directions, respectively. The set of geometrical parameters are defined as follows: $a = 1.3 $ cm, $h_{1} = 0.25 $ cm, $h_{2} = 0.8 $ cm, $w_{1} = 0.1$ cm, $w_{2} = 0.4 $ cm, $w_{3} = 0.55 $ cm,  $g_{1} = 0.63 $ cm, $g_{1} = 0.45 $ cm, $\alpha = 45^{\circ} $ and $\theta  = 70^{\circ} $.}
\end{figure}

The subject of artificially engineered materials has certainly become a highly relevant and intriguing topic in modern-day discourse \cite{wong2017optical}. An important mechanism responsible for the manifestation of numerous physical phenomena is the symmetry breaking that occurs within the mirror-image unit cell containing chiral structures \cite{PhysRevLett.97.167401, hentschel2017chiral,wu2019mechanical,nieves2018vibrations}. The chiral effect, mostly exhibited in biological molecules, embodies an intrinsic characteristic that has a profound impact on the biological activity of conventional chemicals \cite{naaman2019chiral, green2016chiral}. Consequently, the latter constitutes a pivotal determinant in molecule recognition via the chiral-induced spin selectivity process \cite{zhang2012chiral}. However, the advent of plasmonic chiroptical materials has engendered a discernment of a wide spectrum of alluring applications, encompassing circular dichroism, wherein left and right-handed light undergo disparate absorption, circular birefringence, which entails the rotation of the plane of light oscillation, and the emergence of quasi-dark states in the continuum \cite{fan2010plasmonic, schreiber2013chiral, pages2002photoinduced, wu2021observation, mur2014chiral}. In this vein, Yonghao et \textit{al}. demonstrate a strong optical activity in  dual-layered twisted arcs metamaterial with a polarization rotatory of about $305^{\circ}$ \cite{cui2014giant}. The integration of the chiral effect into structural mechanics has recently resulted in significant strides in expanding our knowledge of physical phenomena at both micro and macro scales \cite{qi2021advanced}. One notable phenomenon in structural mechanics is the twist effect, which arises when an externally applied axial deformation interacts with a unit cell possessing ingeniously integrated chirality \cite{frenzel2017three, fernandez2019new}. Another intriguing consequence of the chiral effect is the auxetic effect, denoting a negative Poisson ratio \cite{dudek2022micro,lakes1987foam}. The incorporation of chirality into the unit cell confers upon the studied structure supplementary degrees of freedom, surpassing those that are inherently linked to Cauchy elasticity in systems endowed with an inversion center of symmetry \cite{eringen1999theory, lakes1982noncentrosymmetry}. These supplementary degrees of freedom, namely local rotation and the corresponding couple stress \cite{nowacki1986theory}, in conjunction with local translation and force stress, are imperative for instigating acoustic activity within the system \cite{carta2019wave}. When a linearly polarized elastic wave propagates through an acoustically active medium, it undergoes a transformation in its polarization state, resulting in a circular polarization profile as it spreads \cite{frenzel2019ultrasound}. This transformation of polarization state is contingent upon the handedness of the medium and triggers the wave's polarization state to transition from a linearly polarized state to an orthogonally polarized state. In 1968, Portigal et \textit{al}. hinted at the phenomenon under consideration, proposing that a crystal exhibiting optical activity also manifests acoustic activity, resulting in a rotation of the polarization plane \cite{portigal1968acoustical}. This rotatory effect of polarization states provides adjustable control of one of the intrinsic wave characteristics, effectively providing an additional degree of freedom that can be likened to circular birefringence in acoustic or mechanical waves. Moreover, it has been evidenced that chiral substances exhibit a selective affinity towards sound waves with a particular chirality, which is predicated upon the sound's ability to convey Orbital Angular Momentum (OAM) in the form of vortices \cite{tong2023acoustic,tong2022acoustic}. However, OAM can also be attained in centrosymmetric structures, such as in planar layer resonators and planar arrays of electroacoustic transducers \cite{jiang2016convert, esfahlani2017generation}. In this study, we consider a 3D chiral butterfly meta-structure that has the ability to transform a longitudinally polarized wave into a transversely polarized wave. This conversion is enabled by the coupling of an additional degree of freedom arising from the handedness effect with the compressional wave, which presents an avenue for converting compressional waves to transverse waves and vice versa. We employ an eigenvalue solver based on Finite Element Analysis (FEA) to assess the energy momentum diagram, the total displacement distribution, and the polarization state of the transmitted and reflected waves upon encountering a single homogeneous block, achiral and chiral butterfly unit cells. Additionally, we utilize a classical mass-spring model to theoretically describe the dispersion mode associated to the conversion effect that arises from the coupling of longitudinal and rotational behavior within the chiral unit cell. The geometric design we consider is a 3D butterfly metamaterial with a lattice constant of $a$ and wings angled at $\alpha = 45^{\circ}$. The schematic designs of the non-centrosymmetric and centrosymmetric structures are depicted in [Fig.~\ref{fig:01}(a)] and [Fig.~\ref{fig:01}(b)], respectively.

\begin{figure}[!htbp]
\centering
\includegraphics[width= 8 cm, height = 5.25 cm]{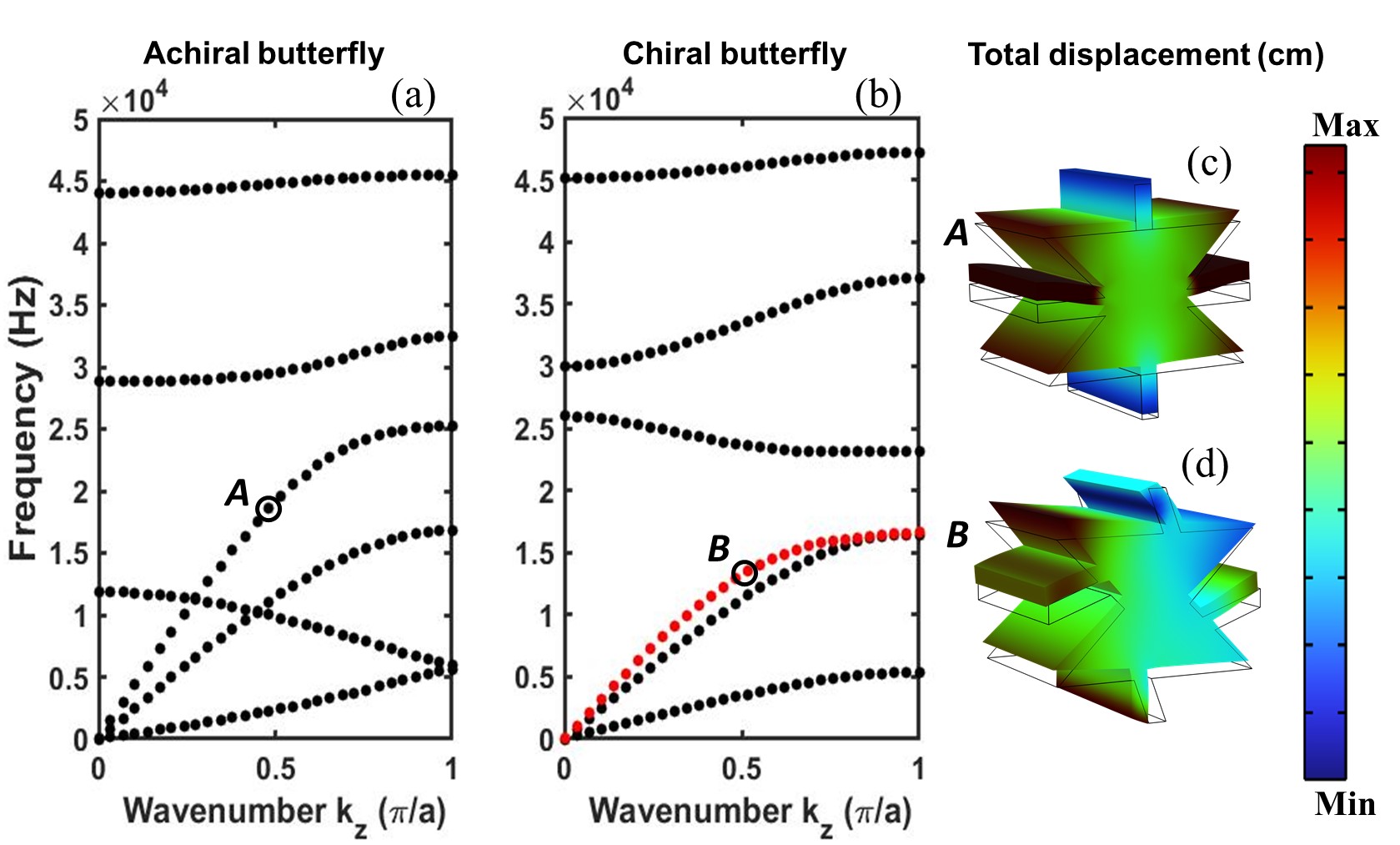}
\caption{\label{fig:02} The energy momentum diagrams for an achiral and chiral butterfly meta-atoms are presented in (a) and (b), respectively. Both meta-atoms have the same spatial period of $a_{x} = a_{y} = a_{z} = a$ along the all directions, with a wavenumber $k_z$ within the range of $[0, \pi/a]$. In (c) and (d), the entire field distribution is illustrated when the wavenumber $k_z$ equals 0.5 for the achiral and chiral meta-atoms, respectively.}
\end{figure}
In the achiral case illustrated in [Fig.~\ref{fig:01}](d), the top-down and left-right connections are precisely orthogonal to the long arms of the connections, with a lateral length of $h_{1}$, and a width of $w_{1}$. Meanwhile, in the chiral case, all the connections undergo a rotation of $\theta = 70^{\circ}$, as shown in [Fig.~\ref{fig:01}](c). All other geometrical parameters can be found in [Fig.~\ref{fig:01}]. Here, we consider the elastic wave propagating along the $[ 0 0 1]$ crystallographic direction with wavenumber $k_{z}$, where $\vec{k}=\left\{ 0, 0, k_{z} \right\}$. The material has a density of $\rho = 1140 $ kg/m$^{3}$, a Poisson ratio of $\nu = 0.4$ and a young modulus of $ E = 4.2 $ GPa.  [Fig.~\ref{fig:02}](a) and [Fig.~\ref{fig:02}](b) delineate the band structure of achiral and chiral meta-atoms, respectively, elucidating all viable modes that may propagate along the $\Gamma Z$ direction. Notably, the total displacement portrayed in [Fig.~\ref{fig:02}](c) and [Fig.~\ref{fig:02}](d) depict the normal longitudinal and rotation coupled-longitudinal motions exhibited by the achiral and chiral butterfly meta-structures, respectively.

\section{\label{sec:level2}Mathematical formulation}

\begin{figure}[!ht]
\centering
\includegraphics[width= 8.6 cm]{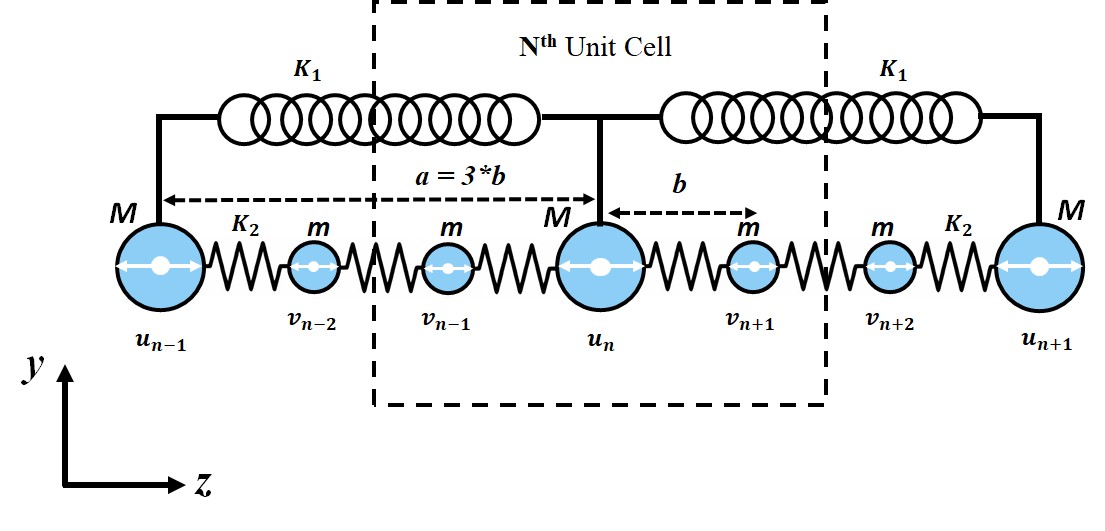}
\caption{\label{fig:03} In the mass-spring model, the sizeable masses $M$ denoted by the large blue dots, are connected to their closest neighbors by a spring featuring a stiffness coefficient of $K_{1}$. This arrangement is characterized by a periodicity of $a=3 b$, which serves as the fundamental unit cell. The successive large masses are joined to two small masses through a spring whose Hook's constant is $K_{2}$ separated by a distance $b$. The incorporation of the supplementary masses within this configuration, pursuant to the microstretch theory, facilitates the assessment of micro-rotational motion.}
\end{figure}

The classical mass-spring model is employed in theory to describe the interplay between changes in length and angle in a 3D chiral butterfly structure. The microstretch continuum is utilized to elucidate the deformational properties of materials by delineating two distinct forms of strain, namely classical and microstretch \cite{eringen1972theory, eringen1990theory}. The former evaluates modifications in the material's length and angle at the macroscopic level, while the latter gauges changes at the microstructural level \cite{eringen2012microcontinuum}. In order to clarify the phenomenon of dilatation within a microstretch medium, a mass-spring model is adopted, which comprises two masses denoted as $M$ and $m$, having unequal weights and linked by two linear springs exhibiting dissimilar stiffness constants labeled as $K_{1}$ and $K_{2}$, respectively. The larger masses $M$ are coupled using two dissimilar linear springs with a spring constant of $K_{1}$ symmetrically placed at a distance of $a$, while the intervening space between two adjacent larger masses $M$ is uniformly divided among two smaller masses $m$. These smaller masses are interconnected with each other and with the larger masses through linear springs having a spring constant of $K_{2}$, as depicted in [Fig.~\ref{fig:03}].

Newton's second law of dynamics expounded the total forces acting on the $n^{th}$ mass $M$ and the two nearest neighbors masses $m$, which can be expressed as follows \cite{eringen2012microcontinuum}:
\begin{equation}\label{eq:1}
\small
\begin{aligned}
    M\frac{\partial^{2} u_{n}}{\partial t^{2}} = K_{1}(u_{n+1} + u_{n-1} -2 u_{n}) + K_{2}(v_{n+1} + v_{n-1} -2u_{n})
    \end{aligned}
\end{equation}

\begin{equation}\label{eq:2}
\begin{aligned}
    m\frac{\partial^{2} v_{n+1}}{\partial t^{2}} = K_{2}(u_{n} + v_{n+2} -2 v_{n+1}) 
\end{aligned}
\end{equation}
\begin{equation}\label{eq:3}
\begin{aligned}
    m\frac{\partial^{2} v_{n-1}}{\partial t^{2}} = K_{2}(u_{n} + v_{n-2} -2 v_{n-1}) 
    \end{aligned}
\end{equation}
In order to account for the significant influence of microrotational motion, Eq.~(\ref{eq:2}) and  Eq.~(\ref{eq:3}), which govern the motion of the two small masses situated on both the left $v_{n-1}$ and right $v_{n+1}$ sides, are combined as shown in the following equation:
\begin{equation}\label{eq:4}
\small
\begin{aligned}
    m\frac{\partial^{2} (v_{n+1} + v_{n-1})}{\partial t^{2}} = K_{2}(2u_{n} + v_{n+2} +v_{n-2}-2 v_{n+1} -2 v_{n-1}) 
    \end{aligned}
\end{equation}
By considering harmonic solutions to  Eq.~(\ref{eq:1}) and  Eq.~(\ref{eq:4}), the periodic conditions can be formulated as follows \cite{lee1973waves, michel1999effective}: $u_{n\pm m} =e^{i(\pm mkb -\omega t )}u_{0}$ and $v_{n\pm m} =e^{i(\pm mkb -\omega t )}v_{0}$, where $k$ and $b = a/3$ are the irreducible wavenumber throughout the $\Gamma Z$ direction and the period between the smallest masses denoted by $m$, respectively. The substitution of the Floquet Block solutions into the governing,  Eq.~(\ref{eq:1}) and  Eq.~(\ref{eq:4}), yields the following relationship:

\begin{equation}\label{eq:5}
 \begin{bmatrix} 
(M \omega^{2} + A )& B \\
C & (m\omega^{2} + D )
\end{bmatrix}
\begin{bmatrix} 
           u_{0} \\
           v_{0}          
\end{bmatrix}=0
\end{equation}
Where the coefficients presented in Eq.~(\ref{eq:5}) A, B, C, and D, are expressed as follows:
\begin{equation}\label{eq:6}
\centering
\left\{\begin{aligned}
A &= K_{1}(cos(3kb) -1)-2K_{2} \\
B &= 2K_{2}cos(kb) \\ 
C &= K_{2}  \\
D &= K_{2}cos(2kb)- 2K_{2}cos(kb) \\
\end{aligned}\right.
\end{equation}
The dispersion relation can be determined by computing the determinant of the system of equations presented in Eq.~(\ref{eq:5}), which results in the following fourth-order equation:
\begin{equation}\label{eq:7}
\begin{aligned}
 \omega^{4} + \underbrace{\frac{(MD+Am)}{Mm}}_{A^{'}} \omega^{2}+ \underbrace{\frac{(AD-BC)}{Mm}}_{B^{'}} = 0
 \end{aligned}
\end{equation}
The roots of Eq.~(\ref{eq:7}) can be expressed as follows:
\begin{equation}\label{eq:8}
\begin{aligned}
\omega_{\pm}= \sqrt{\frac{-A^{'} \pm \sqrt{A^{'2} -4B^{'2}}}{2}}
\end{aligned}
\end{equation}
 \begin{figure}[!htbp]
\centering
\includegraphics[width= 7.2 cm]{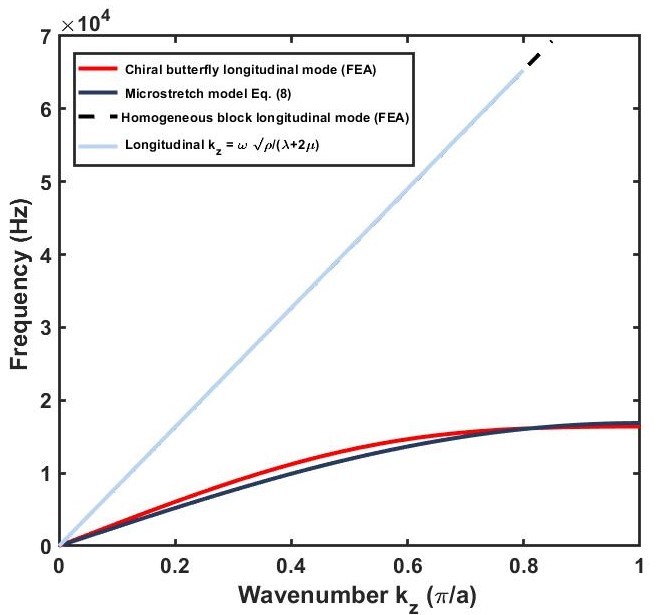}
\caption{\label{fig:04} The diagram showcases the energy-momentum characteristics of both coupled and uncoupled compressional motion, with a particular emphasis on the coupled longitudinal mode demonstrated in the chiral butterfly structure. The curves in red and dark purple correspond to the FEA simulation and microstretch theoretical results, respectively, while those in black and cyan exhibit the dispersion behavior of normal longitudinal waves in a homogeneous 3D block with a spatial period of $a = 1.3$ cm, as predicted by both simulation and theoretical models.}
\end{figure}

[Fig.~\ref{fig:04}] illustrates a comparison between the numerical and theoretical models of the band structure for pure and coupled longitudinal waves observed in both homogeneous blocks and 3D chiral butterfly structures, using the microstretch model with a specific set of parameters obtained through the numerical computation. The implementation of the microstretch model in solids provides an approximate prediction of propagation modes, including the micro-rotational wave. The set of microstretch parameters are presented as follows: $M = 3.3 \times 10^{-3}$ Kg, $m = 1 \times 10^{-5}$ Kg, $a = 1.3 \times 10^{-2}$ m, $ b = a/3$ m, $K_{1} = 2.35 \times 10^{5}$ N/m, $K_{2} = 0.1 \times 10^{1}$ N/m.

\section{\label{sec:level3}Results and discussion}

In this section, we proffer a numerical simulation premised on Finite Element Analysis (FEA) as a means to investigate the capacity of chiral butterfly meta-structures to master the transformation of longitudinal waves to both horizontal and vertical shear waves.
 
\begin{figure}[!ht]
\centering
\includegraphics[width= 8.6 cm, height = 6.5 cm]{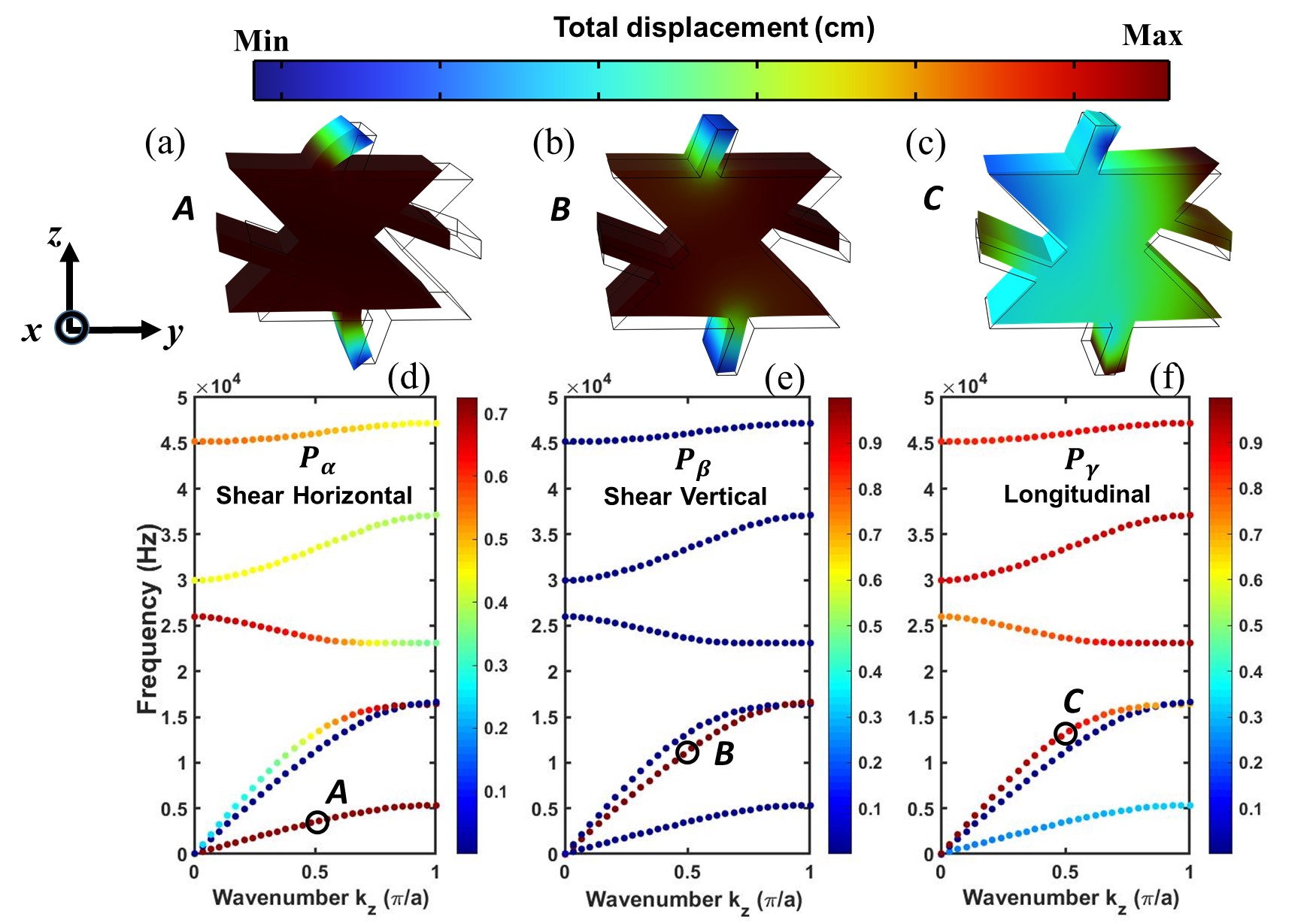}
\caption{\label{fig:05} Panels (a-c) delineate the total displacement behaviors of shear horizontal, shear vertical, and rotational-coupled longitudinal motion at the points  $A$, $B$ and $C$ indicated by the circle in each respective polarization. Panels (d-f) portray the polarization state of the primary triad of modes; namely, shear horizontal, shear vertical, and longitudinal polarizations, respectively.}
\end{figure}
 Our approach employs an eigenvalue problem to scrutinize the polarization states of each mode within the energy momentum diagram for the chiral butterfly structure as illustrated in [Fig.~\ref{fig:05}](d), [Fig.~\ref{fig:05}](e) and [Fig.~\ref{fig:05}](f).

\begin{figure}[!ht]
\centering
\includegraphics[width= 8.6 cm]{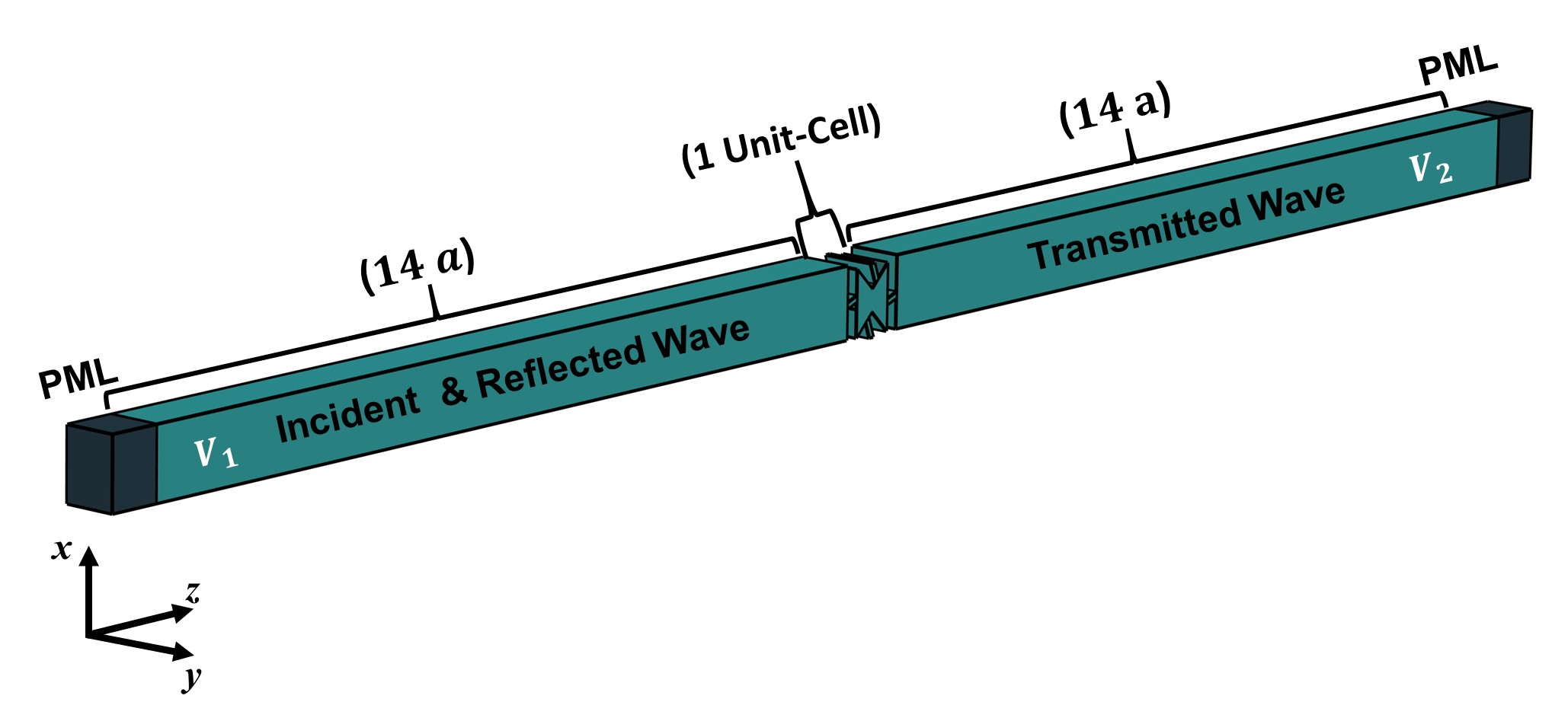}
\caption{\label{fig:06} The three-dimensional schematic representation used for evaluating the polarization state of both transmitted and reflected waves upon interacting with one unit-cell of period $a = 1.3$ cm.}
\end{figure}

\begin{figure}[!ht]
\centering
\includegraphics[width= 8.6 cm]{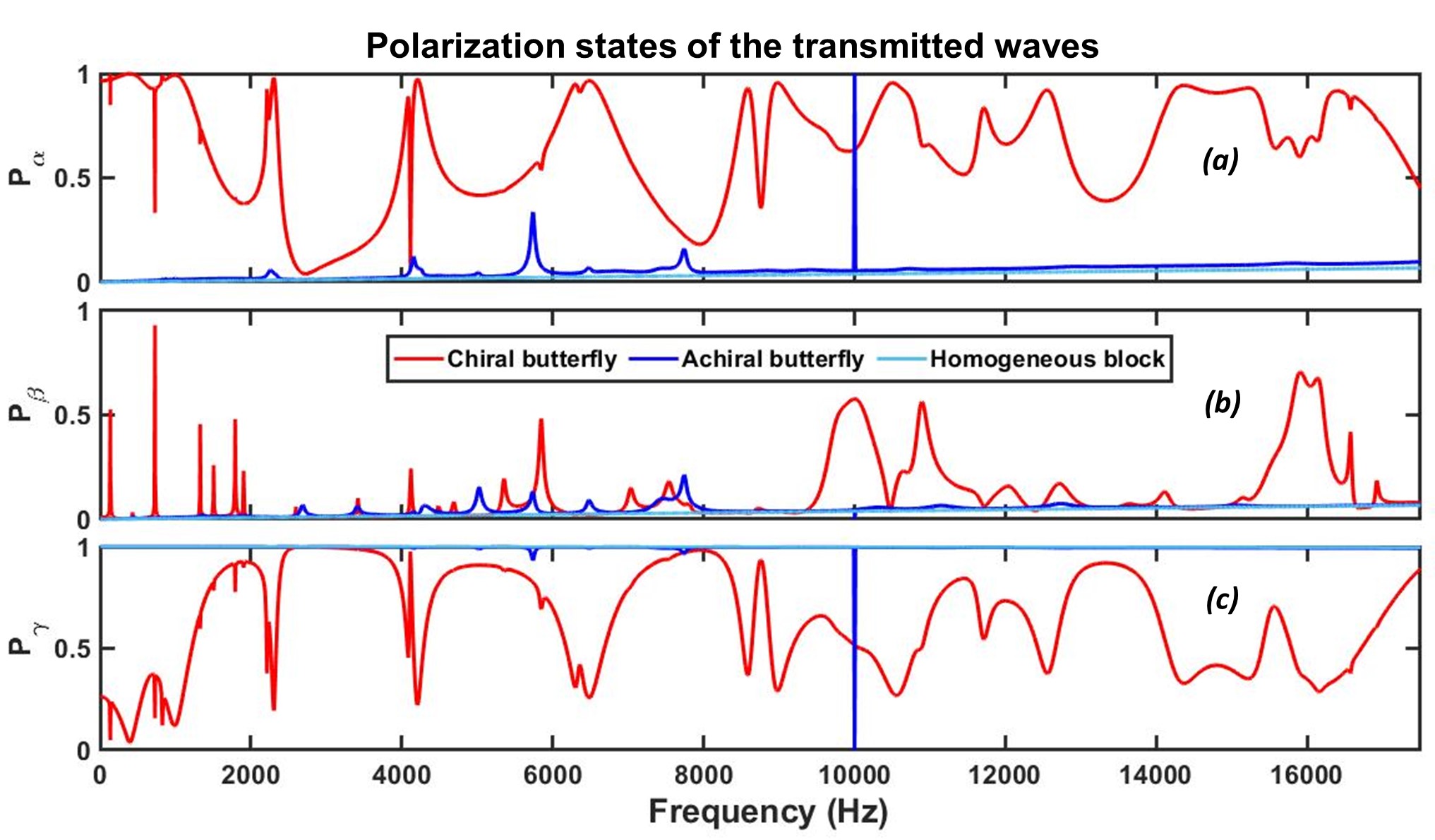}
\caption{\label{fig:07} The diagrams (a), (b), and (c) depict the polarization state of the transmitted elastic wave following interaction with a homogeneous block, an achiral butterfly meta-atom, and a chiral butterfly meta-atom, respectively.}
\end{figure}

These illustrations provide evidence that the longitudinal mode is capable to transform $P_{\gamma}$ to $P_{\alpha}$ and $P_{\beta}$ polarization states, which results from the interaction of the longitudinal wave with rotational motion. The total displacement field associated with the longitudinal mode distinctly displays the ability to amalgamate compressional motion with an additional degree of freedom, namely rotational motion, as is evident from [Fig.~\ref{fig:05}](c). Moreover, a harmonic analysis is conducted to assess the responses of a homogeneous block, achiral and chiral butterfly meta-atoms to an external longitudinal vibration propagating along the \textit{z}-axis. These unit-cells are enclosed by two large building blocks on both sides, composed of the same material as the unit-cell, while two perfectly matched layers (PML) are implemented at the left and right boundaries to preclude any potential reflection, as presented in [Fig.~\ref{fig:06}].
Employing the analytical formulation that defines the polarization states in accordance with the set of Eq.~(\ref{eq:09}) \cite{achaoui2010polarization}, the polarization state of the wave is determined upon encountering a homogeneous block, an achiral butterfly meta-atom, and a chiral butterfly meta-atom. Moreover, an analogous analytical method is utilized to evaluate the polarization state of the reflected waves at the periphery of each of these scrutinized unit cells.

\begin{equation*}
P_{\alpha}=  \frac{\iiint_{V_{i}} \sqrt{|u_{\alpha}|^{2}} \,dr}{\iiint_{V_{i}} \sqrt{|u_{\alpha}|^{2}+|u_{\beta}|^{2}+|u_{\gamma}|^{2}} \,dr} \end{equation*}
\begin{equation}\label{eq:09}
P_{\beta}=  \frac{\iiint_{V_{i}} \sqrt{|u_{\beta}|^{2}} \,dr}{\iiint_{V_{i}} \sqrt{|u_{\alpha}|^{2}+|u_{\beta}|^{2}+|u_{\gamma}|^{2}} \,dr} 
\end{equation}
\begin{equation*}
P_{\gamma}=  \frac{\iiint_{V_{i}} \sqrt{|u_{\gamma}|^{2}} \,dr}{\iiint_{V_{i}} \sqrt{|u_{\alpha}|^{2}+|u_{\beta}|^{2}+|u_{\gamma}|^{2}} \,dr} \end{equation*}
 Eq.~(\ref{eq:09}) utilizes subscripts denoted by $\alpha$, $\beta$ and $\gamma$, respectively representing the polarization and displacement components along the Cartesian coordinates y, x, and z. The parameters $V_{i}$ represent the volumes of the left and right sides of the unit cell, where $i = \left\{ 1,2\right\}$, as indicated in [Fig.~\ref{fig:06}]. By exciting our structure from the left side, as depicted in [Fig.~\ref{fig:06}], with a normalized longitudinal vibration of $1$ and a maximum frequency of $1.8 \times 10^{4}$ Hz, propagating along the \textit{z}-axis, the polarization states of the transmitted waves in a homogeneous block, achiral and chiral meta-atoms, are demonstrated in [Fig.~\ref{fig:07}](a), [Fig.~\ref{fig:07}](b) and [Fig.~\ref{fig:07}](c), respectively. The homogeneous block and the achiral butterfly maintain the same polarization state as the incident wave $P_{\gamma}$, as indicated by the cyan and blue colors, respectively. Conversely, the chiral structure can convert a purely longitudinal polarized wave to both horizontal and vertical shear waves, as represented by the red color. 

\begin{figure}[!ht]
\centering
\includegraphics[width= 8.6 cm]{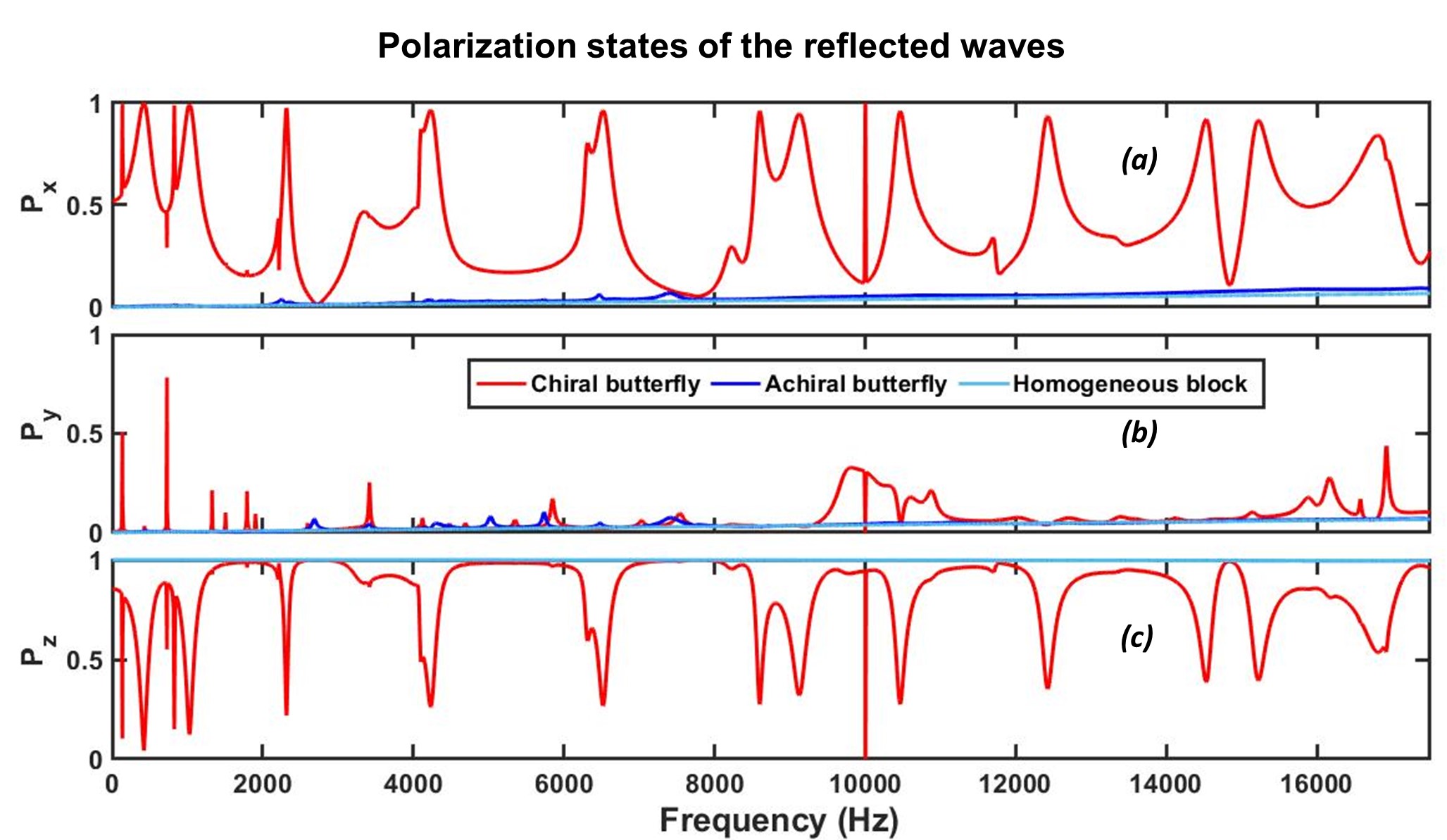}
\caption{\label{fig:08}The illustrations labeled (a), (b), and (c) portray the polarization state of the elastic wave upon reflection at the interface of a homogeneous cube, an achiral butterfly meta-atom, and a chiral butterfly meta-atom, correspondingly.}
\end{figure}

Similarly, the polarization states of reflected waves from the homogeneous block, achiral butterfly meta-structure, and chiral butterfly meta-structures are illustrated in [Fig.~\ref{fig:08}](a), [Fig.~\ref{fig:08}](b) and [Fig.~\ref{fig:08}](c), respectively. 
Furthermore, we evidence that the reflected waves undergo a transformation from a purely longitudinal polarization, $P_{\gamma}$, to alternative polarization states, $P_{\alpha}$ and $P_{\beta}$, upon their interaction with the interface of the chiral butterfly meta-atom. 
\begin{figure}[!h]
\centering
\includegraphics[width= 8.6 cm]{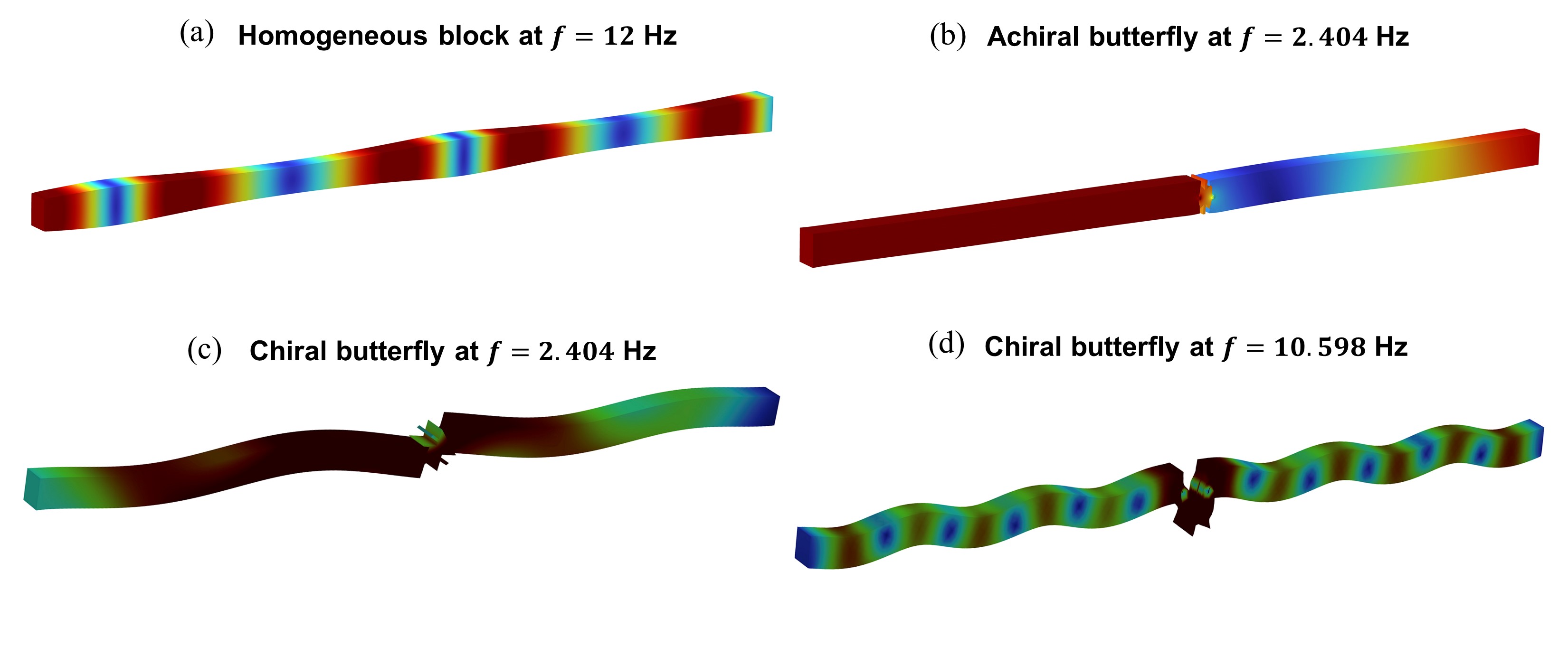}
\caption{\label{fig:09} Panels (a) and (b) present the total displacement field of a homogeneous block and an achiral butterfly unit-cell, respectively, at the frequencies of $12$ kHz and $2.404$ kHz. In contrast, panels (c) and (d) display the total displacement field of the chiral metamaterial unit-cell at the frequencies of $2.404$ kHz and $10.598$ kHz, respectively.}
\end{figure}
Conversely, in the case of the homogeneous block and achiral unit-cells, the polarization states of the reflected waves remain unaltered. The 3D distribution of the total field displacement for the homogeneous, achiral, and chiral butterfly meta-structures is presented in [Fig.~\ref{fig:09}](a-d). In the homogeneous case, the polarization state of both transmitted and reflected waves remains unaltered from the incident wave, denoted as $P_{\gamma}$, as depicted in [Figure~\ref{fig:09}(a)] at the frequency of $12$ kHz. The achiral structure's total displacement field also preserves the same polarization state as the incident excitation throughout the entire spectral range, as exemplified at the frequency of $2.404$ kHz in [Figure~\ref{fig:09}(b)]. In contrast, in [Figures~\ref{fig:09}(c)] and [Figures~\ref{fig:09}(d)], the chiral case manifests the ability to convert both transmitted and reflected waves from longitudinal polarization to shear polarization at the frequencies of $2.404$ kHz and $10.598$ kHz, respectively.
\begin{figure}[!htbp]
\centering
\includegraphics[width= 8.6 cm]{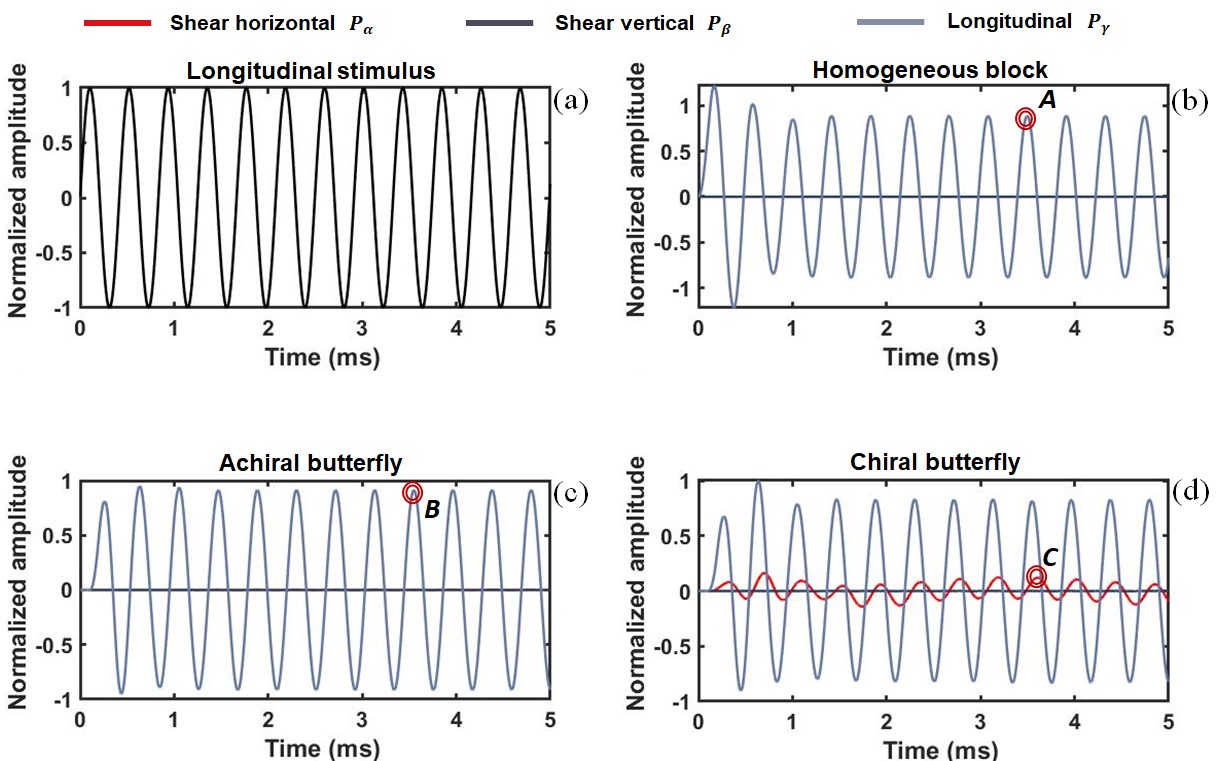}
\caption{\label{fig:10} In panel (a), a longitudinal sinusoidal excitation source with a normalized magnitude of 1 and a frequency of $2.404$ kHz, directed along the z-direction, as indicated by the black curve. Panels (b-d) portray the polarization states of a time-harmonic stimulus after encountering a homogeneous block, as well as achiral and chiral meta-structures, respectively. The fundamental polarizations of the transmitted wave, specifically shear horizontal $P_\alpha$, shear vertical $P_\beta$, and longitudinal $P_\gamma$ for all cases, are delineated in red, blue, and pink, respectively.}
\end{figure}
\begin{figure}[!htbp]
\centering
\includegraphics[width= 8.6 cm]{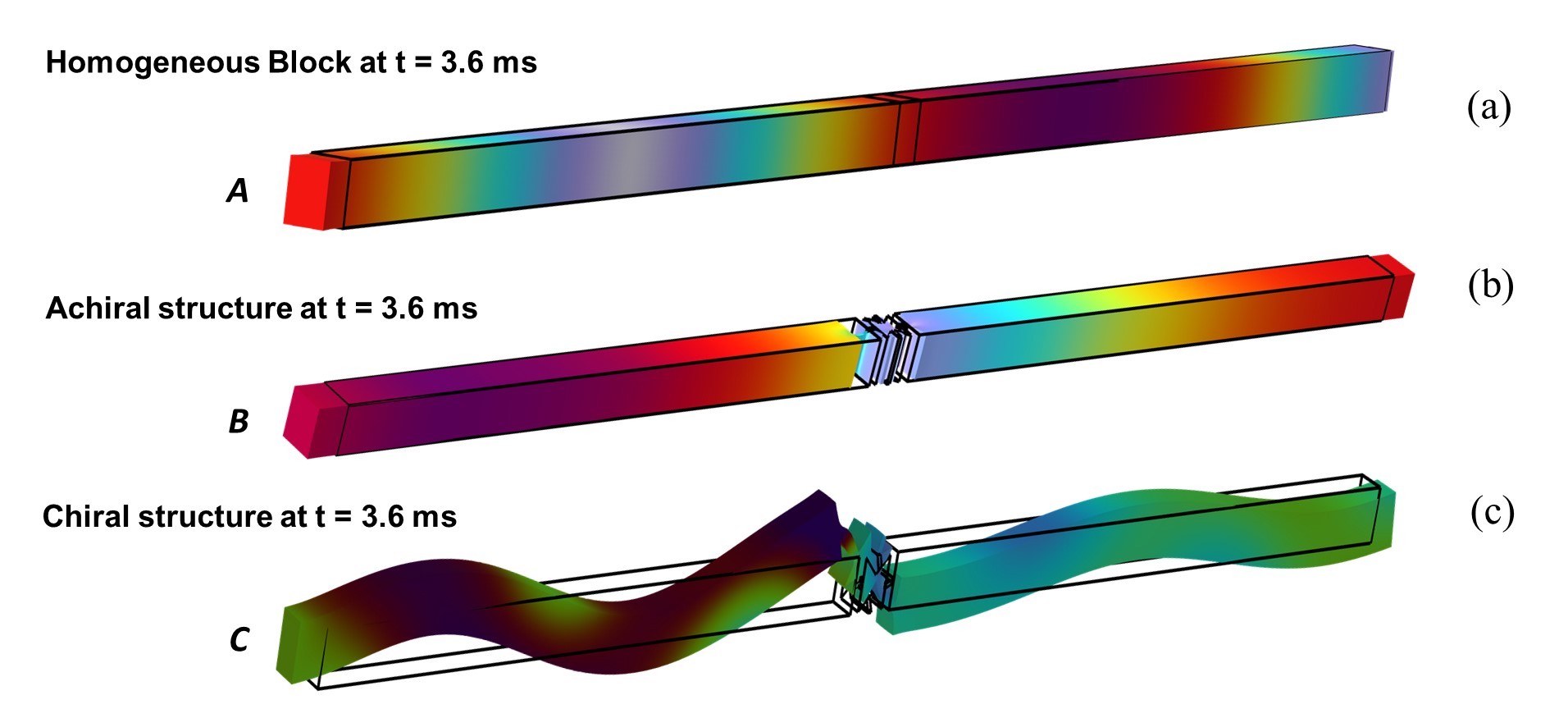}
\caption{\label{fig:11} Panels (a-c) depict the total displacement field at a time step of $t = 3.6$ ms, as denoted by the points \textit{A}, \textit{B} and \textit{C} in [Figures~\ref{fig:10}], for the homogeneous, achiral, and chiral butterfly meta-atoms, respectively.}
\end{figure}
 Additionally, a time-harmonic vibration with a sinusoidal waveform was employed to explore the polarization states of transmitted waves through a homogeneous block, achiral and chiral butterfly meta-atoms. The vibration, directed along the z-axis, possesses a normalized magnitude and a frequency of $2.404$ kHz, with its duration varying from $0$ to $5$ ms, as indicated by the black curve [Figures~\ref{fig:10}(a)]. The findings indicated that the polarization states of transmitted waves remained unchanged when the homogeneous block and achiral unit-cells were stimulated by the incident vibration. This outcome is demonstrated by the pink curves in [Figures~\ref{fig:10}(b)]and [Figures~\ref{fig:10}(c)], where the output showed only the $P_\gamma$ component, while the $P_\alpha$ and $P_\beta$ components remained at zero, as depicted by the red and blue curves, respectively. On the other hand, the chiral butterfly structure converted the incident compressional vibration into a shear one, as observed in [Figures~\ref{fig:10}(d)],  where the $P_\alpha$ component was present in the output signal. Finally, [Figures~\ref{fig:11}(a-c)] displayed the overall displacement field at a specific time of $3.6$ ms for all cases, revealing a remarkable degree of consistency between the frequency and time-harmonic investigations carried out in this study.

\section{\label{sec:level5}Conclusion}

Within the scope of this study, we present a compelling analysis of 3D butterfly structures for facilitating polarization state conversion in solid materials. Our investigation focuses on the mechanical properties of achiral and chiral butterfly structures, where we find that the achiral butterfly, like homogeneous mediums, lack the ability to rotate the polarization plane due to their limited degrees of freedom. Conversely, the incorporation of chiral effects endows the normal butterfly meta-atom with an additional degree of freedom, enabling rotational motion and conservation of the polarization state in both time and frequency domains. These findings demonstrate that this chiral incorporation significantly enables the structure's conversion ability, generating a hybrid mode through compressional and rotational movement. Moreover, we observe effective conversion of transmitted and reflected waves in chiral butterfly structures, confirmed by time harmonic results. Notably, this conversion leads to the transformation of longitudinal waves into corresponding orthogonal polarization, i.e., shear waves. Finally, we highlight that the inverse process of converting incoming shear waves into longitudinal ones is also feasible, and vice versa.

\begin{acknowledgments}
This work was supported by the UTT Project Stratégique NanoSPR (OPE-2022-0293), the Graduate School (Ecole Universitaire de Recherche) “NANOPHOT” (ANR-18-EURE-0013), PHC PROCORE-Campus France/Hong Kong Joint Research Scheme (No. 44683Q) and AAP1-LABEX SigmaPIX 2021. We thank also the EIPHI Graduate School of UBFC [grant number ANR-17-EURE-0002], the French  Investissements d'Avenir program, in part by the ANR PNanoBot (ANR-21-CE33-0015) and ANR OPTOBOTS project (ANR-21-CE33-0003). 
\end{acknowledgments}
\vspace{0.5 cm}
\textbf{Conflict of interest statement:} The authors declare that they do not have any competing interests in the manuscript.\\


\end{document}